\documentclass[10pt]{emulateapj}
\usepackage{amssymb}
\usepackage{pslatex}
\usepackage{graphicx}
\usepackage{psfig}

\begin{document}

\title{Alfv\'en Wave Reflection and Turbulent Heating in the Solar Wind from 1~Solar Radius to 1~AU:  an Analytical Treatment}

\author{Benjamin D. G. Chandran\altaffilmark{1} \&  Joseph V. Hollweg\altaffilmark{1}}

\altaffiltext{1}{Space Science Center and Department of Physics,
       University of New Hampshire, Durham, NH; benjamin.chandran@unh.edu, joe.hollweg@unh.edu}

\begin{abstract}
We study the propagation, reflection, and turbulent dissipation of Alfv\'en waves in coronal holes and the solar wind. We start with the Heinemann-Olbert equations, which describe non-compressive magnetohydrodynamic fluctuations in an inhomogeneous medium with a background flow parallel to the background magnetic field. Following the approach of Dmitruk et al., we model the nonlinear terms in these equations using a simple phenomenology for 
the cascade and dissipation of wave energy, and assume that there is much more energy in waves propagating away from the Sun than waves propagating towards the Sun. We then solve the equations analytically for waves with periods of hours and longer to obtain expressions for the wave amplitudes and turbulent heating rate as a function of heliocentric distance. We also develop a second approximate model that includes waves with periods of roughly one minute to one hour, which undergo less reflection than the longer-period waves, and compare our models to observations. Our models generalize the phenomenological model of Dmitruk et al.\ by accounting for the solar wind velocity, so that the turbulent heating rate can be evaluated from the coronal base out past the Alfv\'en critical point - that is, throughout the region in which most of the heating and acceleration occurs. The simple analytical expressions that we obtain can be used to incorporate Alfv\'en-wave reflection and turbulent heating into fluid models of the solar wind.
\end{abstract}


\maketitle

\section{Introduction}
\label{intro}

Beginning with the discovery of large-amplitude Alfv\'en waves in the
interplanetary medium four decades ago (Belcher, Davis, \& Smith 1969;
Belcher \& Davis 1971), spacecraft measurements have demonstrated that
turbulent fluctuations in the magnetic field and plasma velocity are
ubiquitous in the solar wind (see, e.g., Tu \& Marsch 1995; Goldstein,
Roberts, \& Matthaeus~1995; Bruno \& Carbone 2005). As in hydrodynamic
turbulence, fluctuation energy in turbulent magnetized plasmas
cascades from large scales to small scales. Upon reaching sufficiently
small scales (typically, of order the proton gyroradius and smaller),
the energy is dissipated, thereby heating the ambient plasma [see
Schekochihin et al.~(2009) for a recent and detailed
discussion]. Turbulent heating of the solar wind is thus inevitable,
and may play an important role in the origin and evolution of the
solar wind (see, e.g., Coleman 1968; Barnes 1981; Hollweg 1983; Tu
et al. 1984; Hollweg 1986; Tu 1987, 1988; Hollweg \& Johnson 1988;
Velli et al. 1989, 1990; Grappin and Velli 1991; Velli 1991;
Grappin et al. 1993; Marsch and Tu 1996; Matthaeus et al. 1999).

The bulk of the fluctuation energy in solar-wind turbulence is at
length scales greatly exceeding the proton gyroradius, is
non-compressive, and to a good approximation can be treated within the
approximation of reduced magnetohydrodynamics (MHD) (Kadomtsev \&
Pogutse~1974; Strauss~1976; Zank \& Matthaeus~1992; Schekochihin et
al~2009). The cascade of energy to smaller scales in reduced MHD
turbulence arises only from interactions between oppositely
propagating Alfv\'en waves. It has thus been known for some time that
the launching of outward-propagating Alfv\'en waves by the Sun is not
sufficient to cause turbulent heating, unless some additional process
generates Alfv\'en waves that propagate towards the Sun (in the
solar-wind frame). Different mechanisms for producing
Sunward-propagating waves have been considered, including the
parametric instability (Galeev \& Oraevskii 1963; Vi\~nas \& Goldstein
1991; Del Zanna, Velli, \& Londrillo 2001) and instabilities driven by
velocity shear (Roberts et al. 1992; Breech et al., 2008). In this
paper, we focus on a third mechanism: the non-WKB reflection of
Alfv\'en waves.

When the non-uniformity of the background plasma is taken into
account, purely outward propagating waves are no longer a solution to
the reduced MHD equations. Instead, outward and inward waves are
coupled through linear terms that are proportional to the gradients in
the Alfv\'en speed (Velli 1993; Hollweg \& Isenberg~2007). This coupling gives rise to wave
reflection. At sufficiently high frequencies, linear Alfv\'en waves
propagating in a hypothetical, steady solar wind are insensitive to
radial variations in the Alfv\'en speed and undergo little
reflection. On the other hand, for wave periods exceeding $\sim
20$~minutes, the transmission coefficient for outward-propagating
Alfv\'en waves launched from the Sun drops significantly below unity
(Velli 1993).  A number of authors (e.g., Velli et al. 1989; Matthaeus
et al. 1999; Dmitruk et al. 2001; Dmitruk \& Matthaeus 2003;
Chandran et al. 2009; Verdini et al. 2009) have discussed the
propagation of Alfv\'enic fluctuations originating at the Sun,
emphasizing the self-consistent generation of reflected waves and the
resulting cascade and dissipation of wave energy. In particular,
Dmitruk et al. (2002, hereafter ``D02'') developed a simple
phenomenological model describing reflections and turbulence
and obtained an analytical approximation for the turbulent heating
rate valid close to the Sun in the limit of strong turbulent
dissipation.  Elements of their approach were incorporated into
detailed solar wind models by Cranmer \& van Ballegooijen (2005, hereafter CvB05),
Cranmer et al. (2007), and Verdini \& Velli (2007, hereafter VV07).

In view of the potential usefulness of D02's analytical approximation,
our purpose here is to extend their approximation out to the Alfv\'en
critical point and beyond, i.e., throughout the region where the
principal solar wind heating and acceleration occur, and to obtain
analytical formulas for the radial profiles of the turbulent heating
rate and wave amplitudes.  In section~\ref{sec:lwfm} we focus on waves
with periods of hours and longer.  In section~\ref{sec:extended}, we
extend our model to account for shorter-period waves.

\section{Low-Wave-Frequency Model}
\label{sec:lwfm} 

We begin with the usual equations of ideal MHD,
\begin{equation}
\frac{\partial \rho}{\partial t} + \nabla \cdot (\rho {\bf v}) = 0,
\label{eq:cont} 
\end{equation} 
\begin{equation}
\rho \left(\frac{\partial}{\partial t} + {\bf v} \cdot \nabla\right) {\bf v} = 
- \nabla p_{\rm tot} + \frac{{\bf B} \cdot \nabla {\bf B}}{4\pi}  - \rho \nabla \Phi,
\label{eq:momentum} 
\end{equation} 
and
\begin{equation}
\frac{\partial {\bf B}}{\partial t} = \nabla \times ({\bf v} \times {\bf B}),
\label{eq:induction} 
\end{equation} 
where $\rho$, ${\bf v}$, and ${\bf B}$ are the mass density, velocity,
and magnetic field, $ \Phi$ is the gravitational potential, $p_{\rm
  tot} = p + B^2/8\pi$ is the total pressure, and $p$ is the plasma
pressure. We set
\begin{equation} 
{\bf v}   =  {\bf U} + \delta {\bf v}
\end{equation}
and
\begin{equation} 
{\bf B}   =  {\bf B}_0 + \delta {\bf B},
\end{equation} 
neglect the Sun's rotation, and take the background flow velocity ${\bf U}$ to be aligned with the background magnetic field~${\bf B}_0$.  We assume that $\delta {\bf v}$ and $\delta{\bf B}$ are perpendicular to ${\bf B}_0$ and  non-compressive. We neglect density fluctuations\footnote{To make this assumption self-consistent, we must also neglect spatial variations in the background density in the directions perpendicular to ${\bf B}_0$.} and take
$\rho$, ${\bf U}$, and ${\bf B}_0$ to be steady-state solutions of  equations~(\ref{eq:cont}) through  (\ref{eq:induction}).
The vector Alfv\'en velocity is given by 
\begin{equation}
{\bf v}_{\rm A} = \frac{{\bf B}_0}{\sqrt{4\pi \rho}},
\label{eq:defva} 
\end{equation} 
and the Els\"asser variables are defined as
\begin{equation}
{\bf z}^\pm = \delta {\bf v} \mp \delta {\bf b},
\label{eq:Elsasser} 
\end{equation} 
where
$\delta {\bf b} = \delta {\bf B}/\sqrt{4\pi \rho}$.
Re-writing equations~(\ref{eq:momentum}) and (\ref{eq:induction}) in terms of ${\bf z}^\pm$, we obtain (Velli~1993; VV07)
\[
\frac{\partial {\bf z}^\pm}{\partial t}  
+\left({\bf U} \pm {\bf v}_{\rm A}\right)\cdot \nabla{\bf z}^\pm 
+ {\bf z}^\mp \cdot \nabla \left({\bf U} \mp {\bf v}_A\right)
\]
\begin{equation}
+ \frac{1}{2} \left({\bf z}^- - {\bf z}^+\right) \left(\nabla \cdot{\bf v}_{\rm A} 
\mp \frac{1}{2}\nabla \cdot {\bf U} \right) = - \left( {\bf z}^\mp \cdot \nabla {\bf z}^\pm
+ \frac{\nabla p_{\rm tot}}{\rho}\right).
\label{eq:velli} 
\end{equation} 
The $\nabla p_{\rm tot}$ term cancels the compressible part of the
remaining terms in equation~(\ref{eq:velli}) to maintain the
incompressibility condition~$\nabla \cdot {\bf z}^\pm = 0$.

To proceed further, we assume that the background magnetic field~${\bf B}_0$ possesses a field line that is purely radial. This field line can be at any heliographic latitude.
We then consider a cylindrical coordinate system $(R, \phi, z)$ whose $z$ axis is aligned with this radial field line, and we restrict our analysis to the region close to the $z$ axis, in which we assume that $v_{A\phi} = U_\phi = 0$ and $\partial  U/\partial \phi = \partial v_{\rm A}/\partial \phi = 0$. Near the $z$ axis, the components of ${\bf U}$ and ${\bf v}_{\rm A}$ can be expanded in powers of~$R$,
\begin{eqnarray} 
U_R &  = & a_1(z) R + \dots, 
\label{eq:UR} 
\\
v_{{\rm A}, R} & = & a_2(z) R + \dots
\label{eq:vAR} 
\end{eqnarray}
Keeping just the first terms in these Taylor series,
we find that $\partial U_R/\partial R = U_R/R$, $\partial v_{{\rm A},R}/\partial R = v_{{\rm A},R}/ R$, and  
\begin{equation}
{\bf z}^\mp \cdot \nabla \left({\bf U} \mp {\bf v}_{\rm A}\right)
 = \frac{{\bf z}^\mp (U_R \mp v_{{\rm A},R})}{R}.
\label{eq:approx1} 
\end{equation} 
We define
\begin{equation}
 \frac{\partial}{\partial s} = ({\bf B}_0/B_0) \cdot \nabla  ,
\label{eq:defs} 
\end{equation} 
so that $s$ is distance along a background magnetic field line.  Since $\nabla R = \hat{R} $,
\begin{equation}
\frac{U_R \mp v_{{\rm A},R}}{R} = \frac{(U\mp v_{\rm A})}{R} \;\frac{\partial R(s)}{\partial s},
\label{eq:Rs} 
\end{equation}
where $R(s)$ is to be interpreted as the distance from the
$z$ axis to a given background magnetic field line.

We define vector versions of the Heinemann \& Olbert (1980) variables ${\bf f}$ and ${\bf g}$ as follows:
\begin{eqnarray} 
{\bf z}^+ & = & \frac{{\bf g} \,\eta^{1/4}}{1 + \eta^{1/2}} 
\label{eq:defh} 
\\
{\bf z}^- & = & \frac{{\bf f} \,\eta^{1/4}}{1 - \eta^{1/2}} ,
\label{eq:deff} 
\end{eqnarray} 
where
\begin{equation}
\eta = \rho/\rho_a,
\label{eq:defeta} 
\end{equation} 
and $\rho_a$ is the value of $\rho$ at the Alfv\'en critical point (on the $z$ axis).
Mass conservation requires that $\partial (\rho U/B_0)/\partial s = 0$, and thus $\eta^{1/2} = v_{\rm A}/U$. With the use of equations~(\ref{eq:approx1}) and (\ref{eq:Rs}), equation~(\ref{eq:velli})  can be re-written as (VV07)\footnote{Scalar versions of the linear terms in
  equations~(\ref{eq:Hh}) and (\ref{eq:Hf}) (for just the $\phi$
  components of ${\bf g}$ and ${\bf f}$) were obtained by Heinemann \&
  Olbert (1980), who took the fluctuations to be axisymmetric with
  ${\bf z}^\pm\propto \hat{\phi}$.}
\[
\frac{\partial {\bf g}}{\partial t} + (U + v_{\rm A}) \frac{\partial {\bf g}}{\partial s}
+ (U+v_{\rm A}) {\bf f} \frac{\partial L}{\partial s}  \hspace{3cm} 
\]
\begin{equation}
 =   - \left(\frac{1+\eta^{1/2}}{\eta^{1/4}}\right)
\left({\bf z}^- \cdot \nabla {\bf z}^+ + \frac{\nabla p_{\rm tot}}{\rho}\right)
\label{eq:Hh} 
\end{equation} 
\[
\frac{\partial {\bf f}}{\partial t} + (U - v_{\rm A}) \frac{\partial {\bf f}}{\partial s}
+  (U-v_{\rm A}) {\bf g} \frac{\partial L}{\partial s}   \hspace{3cm} 
\]
\begin{equation} 
 =    \left(\frac{\eta^{1/2} - 1}{\eta^{1/4}}\right)
\left({\bf z}^+ \cdot \nabla {\bf z}^- + \frac{\nabla p_{\rm tot}}{\rho}\right),
\label{eq:Hf} 
\end{equation} 
where 
$L = \ln[R(s)\eta^{1/4}/R_{\rm ref}]$,
and $R_{\rm ref}$ is an arbitrary constant. 
Close to the $z$ axis, $\partial(B_0 R^2)/\partial s = 0$, and $R_{\rm ref}$ can be chosen so that 
\begin{equation}
L = - \frac{1}{2} \ln \left(\frac{v_{\rm A}}{v_{{\rm A}a}}\right),
\label{eq:defL} 
\end{equation} 
where $v_{{\rm A}a}$ is the Alfv\'en speed at the Alfv\'en critical point on the $z$ axis.

Dmitruk et al. (2002) numerically solved equation~(\ref{eq:velli}) in
the limit $U\rightarrow 0$.  They found that the numerical results
could to a large extent be understood in terms of a phenomenological
model in which ${\bf z}^\mp \cdot \nabla {\bf z}^\pm $ is replaced by
$|z^\mp| {\bf z}^\pm / 2\lambda(s)$. We employ the same
phenomenological model [see Chandran et al (2009) for a more detailed discussion], replacing the nonlinear terms ${\bf z}^\mp
\cdot \nabla {\bf z}^\pm + \rho^{-1} \nabla p_{\rm tot}$ in
equations~(\ref{eq:Hh}) and (\ref{eq:Hf}) with $ |z^\mp| {\bf z}^\pm /
2\lambda(s)$.  Equations~(\ref{eq:Hh}) and (\ref{eq:Hf}) then become
\begin{equation} 
\frac{\partial {\bf g}}{\partial t} + (U + v_{\rm A}) \frac{\partial {\bf g}}{\partial s}
+ (U+v_{\rm A}) {\bf f} \frac{\partial L}{\partial s}  
= - \frac{\eta^{1/4}}{2\lambda} \left|\frac{f}{1 - \eta^{1/2}}\right| {\bf g}
\label{eq:Hh2} 
\end{equation} 
and 
\begin{equation} 
\frac{\partial {\bf f}}{\partial t} + (U - v_{\rm A}) \frac{\partial {\bf f}}{\partial s}
+  (U-v_{\rm A}) {\bf g} \frac{\partial L}{\partial s}   =
- \frac{\eta^{1/4}|g|{\bf f}}{2\lambda(1 + \eta^{1/2})} .
\label{eq:Hf2} 
\end{equation} 

Observations (e.g. Belcher and Davis 1971) show that most of the Alfv\'enic
power in the solar wind is at long periods, of the order of hours, as
seen in the spacecraft frame. In this section, we thus restrict our
attention to waves with periods of hours and longer in the frame of the Sun.  We
further assume that the amplitude of the outward propagating waves is
sufficiently large, and $\lambda$ is sufficiently small, that the time
scale for the cascade and dissipation of $z^-$ fluctuations,
$\lambda/z^+$, is shorter than the wave period. In this case, we can
neglect the first and second terms on the left-hand side of
equation~(\ref{eq:Hf2}) and solve equation~(\ref{eq:Hf2}) to obtain
\begin{equation}
{\bf f} = \frac{ 2\lambda (\eta - 1) U}{\eta^{1/4}} \;\frac{\partial L}{\partial s} \;\hat{g},
\label{eq:fsolve} 
\end{equation} 
where $\hat{g} = {\bf g}/g$. We assume that the rapid cascading of $z^-$ energy leads to the inequality
\begin{equation}
z^- \ll z^+,
\label{eq:fapprox} 
\end{equation} 
so that we can neglect the third term on the left-hand side of
equation~(\ref{eq:Hh2}). Taking the dot product of
equation~(\ref{eq:Hh2}) with ${\bf g}$, and then taking the time
average of the resulting equation, we find that
\begin{equation}
\frac{\partial }{\partial s} \langle g^2 \rangle = - 2 \langle g^2 \rangle  \left| \frac{\partial L}{\partial s}\right|,
\label{eq:hdiffeq} 
\end{equation} 
where $\langle \dots \rangle$ denotes a time average. Since we
restrict our analysis to the vicinity of our $z$ axis, $s\simeq r$,
where $r$ is the radial coordinate in spherical coordinates centered
on the Sun.  We take $L$ to be a decreasing function of $r$ near the
Sun, to have a minimum at $r=r_m$ (corresponding to the maximum of
$v_{\rm A}$), and to increase with $r$ at $r>r_m$. We take $r_m$ to be
less than the radius~$r_a$ of the Alfv\'en critical point. Integrating
equation~(\ref{eq:hdiffeq}), we find that
\begin{equation}
\langle g^2 \rangle ^{1/2} = g_a \left(\frac{v_{\rm A}}{v_{{\rm A}a}}\right)^{1/2}
\label{eq:hsolve_out} 
\end{equation} 
for $r> r_m$, where $g_a$ is the rms value of $g$ at $r=r_a$. For $r<r_m$, we find that
\begin{equation}
\langle g^2 \rangle ^{1/2} = g_a \left(\frac{v_{{\rm A}m}^2}{v_{{\rm A}a}v_{\rm A}}\right)^{1/2},
\label{eq:hsolve_in} 
\end{equation} 
where $v_{{\rm A}m}$ is the Alfv\'en speed at $r=r_m$.

The energy density of the Alfv\'enic fluctuations is $\rho [(z^+)^2 + (z^-)^2]/4$. We assume that the energy that is drained from $z^+$ and $z^-$ fluctuations by the nonlinear ``damping'' terms in our model is converted to thermal energy. The turbulent heating rate, for $z^- \ll z^+$, is thus
\begin{equation}
Q = \frac{\rho |z^-| (z^+)^2}{4\lambda}.
\label{eq:defQ} 
\end{equation} 
With the use of equations~(\ref{eq:fsolve}), (\ref{eq:hsolve_out}), and (\ref{eq:hsolve_in}), we find that the average heating rate is
\begin{equation}
Q = \frac{\rho \eta^{1/2} g_a^2 U}{4 (1 + \eta^{1/2}) v_{{\rm A }a}} \; \left|
\frac{dv_{\rm A}}{dr}\right|
\label{eq:Qout} 
\end{equation} 
for $r>r_m$, and
\begin{equation}
Q = \frac{\rho \eta^{1/2} g_a^2 U}{4 (1 + \eta^{1/2})}\left(\frac{v_{{\rm A}m}^2}{v_{{\rm A}a} v_{\rm A}^2}\right) \frac{dv_{\rm A}}{dr}
\label{eq:Qin} 
\end{equation} 
for $r< r_m$.

\begin{figure}[h]
\centerline{\includegraphics[width=8.cm]{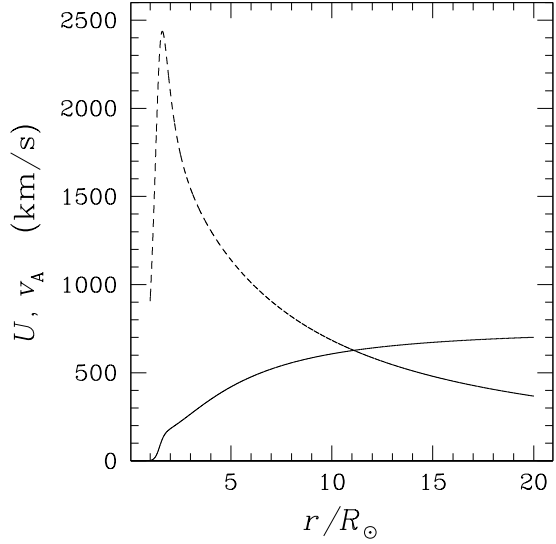}}
\caption{The solar wind velocity (solid line) and Alfv\'en speed
  (dashed line) from Equations~(\ref{eq:n}) through (\ref{eq:defU})
  with $f_{\rm max} = 5$.}\vspace{0.5cm}
\label{fig:swuva}
\end{figure}

To check the validity of some of our approximations and compare to
observations, we consider a model solar wind in which the proton
number density is given by equation~(4) of Feldman et al.~(1997),
which describes coronal holes out to several solar radii, plus an
additional~$r^{-2}$ component:
\begin{equation}
n(r) = \left(\frac{3.23 \times 10^8}{x^{15.6}} + \frac{2.51 \times 10^6}{x^{3.76}} + \frac{1.85 \times 10^5}{x^2}\right) \mbox{ cm}^{-3},
\label{eq:n} 
\end{equation} 
where $x = r/R_{\sun}$ and $R_{\sun} = 1$ solar radius. This leads to $n=4
\mbox{ cm}^{-3}$ at 1~AU.  We take the magnetic field strength to be
(Hollweg \& Isenberg 2002)
\begin{equation}
B_0 = \left[\frac{1.5(f_{\rm max} - 1)}{x^6} + \frac{1.5}{x^2}\right] \mbox{ Gauss},
\label{eq:B0} 
\end{equation} 
where $f_{\rm max}$ is the usual  super-radial expansion factor, which we set equal to~5. The solar wind speed $U$ is given by flux conservation:
\begin{equation}
U = 9.25 \times 10^{12} \;\frac{\tilde{B}}{\tilde{n}} \;\mbox{cm}\;\mbox{s}^{-1} ,
\label{eq:defU} 
\end{equation} 
where $\tilde{B}$ is $B_0$ in Gauss and $\tilde{n}$ is $n$ in units of $\mbox{cm}^{-3}$.
Equation~(\ref{eq:defU}) implies a flow speed of $750\; \mbox{km}\; \mbox{s}^{-1}$ and a
proton flux of $3 \times 10^8\; \mbox{cm}^{-2}\;\mbox{s}^{-1}$ at
1~AU. The Alfv\'en critical point in this model is at $r_a=11.1 R_{\sun}$,
and the maximum of $v_{\rm A}$ is at $r_m= 1.60 R_{\sun}$. The solar wind
speed and Alfv\'en speed are plotted in Figure~\ref{fig:swuva}. 

We set
\begin{equation}
\lambda = \lambda_0 \left[\frac{B(R_{\sun})}{B}\right]^{1/2},
\label{eq:lambda} 
\end{equation} 
where $B(R_{\sun}) = 7.5$~G from equation~(\ref{eq:B0}), and
$\lambda_0$ is a constant.  Previous studies suggest a range of
possible values for~$\lambda_0$.  One choice is to set $\lambda_0$
equal to the mean spacing between photospheric flux tubes.  Following
Spruit (1981), we take this mean spacing to be $R_p \sqrt{\pi
  B_p/B_{\rm avg}}$, where $R_p= 200$~km and $B_p=1500$~G are the
typical radius and magnetic field strength of a photospheric flux
tube, and $B_{\rm avg}$ is the average magnetic field at the coronal
base above the flux-tube merging height. If we set $B_{\rm avg}$ equal
to be the value of $B_0$ in equation~(\ref{eq:B0}) as $x\rightarrow 1$
(which is 7.5~G), then $\lambda_0 = 5000$~km. On the other hand, D02
used $\lambda_0 = 3\times 10^4$~km (comparable to the average size of
supergranules), CvB05 considered values of $\lambda_0$ in the range
$1-30 \times 10^3$~km, and the quantity corresponding to
our~$\lambda_0$ in VV07 was $1.7\times 10^4$~km. We will consider
several values of~$\lambda_0$ in the range spanned by the above
values.

\begin{figure}[t]
\centerline{\includegraphics[width=8.cm]{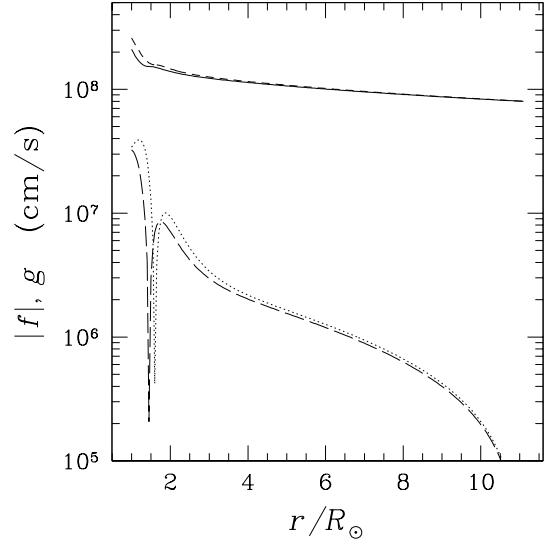}}
\caption{A comparison between our approximate solutions for $|f|$
  (dotted line) and $g$ (short-dashed line) and numerical solutions to
  equations~(\ref{eq:Hh2}) and (\ref{eq:Hf2}) obtained in the limit of
  zero frequency.  The solid line is the numerical solution for
  $g$ and the long-dashed line is the numerical solution for
  $|f|$. For these curves, we have taken $g_{\rm a} = 8.0 \times 10^7 \,\mbox{cm/s}$ and
  $\lambda_0 = 5000$~km.}\vspace{0.5cm}
\label{fig:fgap}
\end{figure}

To check whether we are justified in neglecting the terms that we have
dropped in equations~(\ref{eq:Hh2}) and (\ref{eq:Hf2}), it would be
desirable to compare our approximate solutions to exact,
time-dependent solutions of equations~(\ref{eq:Hh2}) and
(\ref{eq:Hf2}). However, such time-dependent solutions are beyond the
scope of this study.  Instead, we numerically solve the
time-independent versions of equations~(\ref{eq:Hh2}) and
(\ref{eq:Hf2}), taking $f=0$ at the Alfv\'en critical point, setting $\lambda_0 = 5000$~km and $g_a =
8 \times 10^7 \mbox{cm/s}$, and assuming that the directions of ${\bf
  f}$ and ${\bf g}$ are the same and do not change with~$s$. This
steady-state numerical solution is still a useful reference point
since we are considering only long-period oscillations in this
section.  In Figure~\ref{fig:fgap}, we plot the numerical solutions
for $|f|$ and $g$, as well as our approximate analytical solutions
from equations~(\ref{eq:fsolve}), (\ref{eq:hsolve_out}), and
(\ref{eq:hsolve_in}). For $r>2R_{\sun}$, our approximate solutions are
close to the numerical solutions. Closer to the Sun, our analytical
expressions over-estimate the values of $|f|$ and $g$.  In particular,
our approximate value of $g$ is 23\% larger than the numerical solution
at $r=R_{\sun}$. If $\lambda_0$ is increased to $10^4$~km ($1.5\times 10^4$~km), the
approximate solution is 43\% (57\%) larger than the numerical solution
at $r=R_{\sun}$, although the approximate solutions remain reasonably
accurate at $r>2R_{\sun}$. Increasing $\lambda_0$ reduces the nonlinear
terms in equations~(\ref{eq:Hh2}) and (\ref{eq:Hf2}) and degrades the
accuracy of our approximations.

\begin{figure}[t]
\centerline{\includegraphics[width=8.cm]{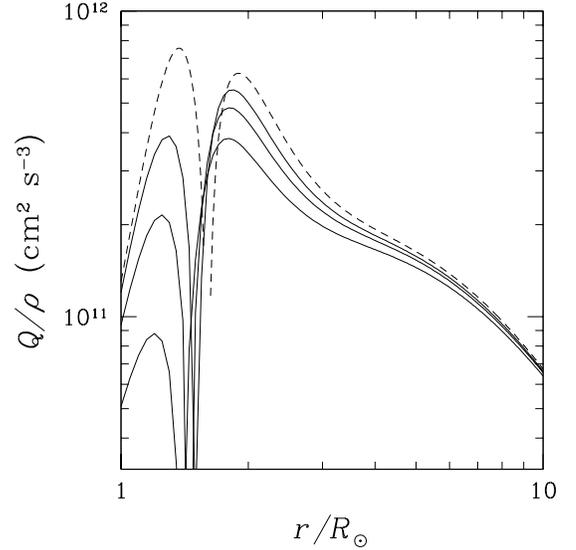}}
\caption{The dashed line is the analytical approximation of the
  heating rate per unit mass~$Q/\rho$ from equations~(\ref{eq:Qout})
  and (\ref{eq:Qin}), with $g_{\rm a} = 8.0 \times 10^7
  \,\mbox{cm/s}$. The solid lines are the total heating rate $Q_{\rm total}
  = \rho [(z^+)^2 |z^-| + (z^-)^2 |z^+|]/4\lambda$ divided by~$\rho$
  in numerical solutions of the time-independent versions of
  equations~(\ref{eq:Hh2}) and (\ref{eq:Hf2}), with $g_{\rm a} = 8.0
  \times 10^7 \,\mbox{cm/s}$ and with three different values
  of~$\lambda_0$: 2500~km, 5000~km, and $10^4$~km (going from the top
  curve to the bottom curve). As $\lambda_0$ increases, our
  approximation becomes increasingly inaccurate.}\vspace{0.5cm}
\label{fig:swh_comp}
\end{figure}

The loss of accuracy in our analytical approximations as $\lambda_0$
increases is further illustrated in Figure~\ref{fig:swh_comp}, which
is analogous to Figure~5 of D02.  The dashed line in this figure shows
$Q/\rho$ from equations~(\ref{eq:Qout}) and (\ref{eq:Qin}). The three
solid lines plot the total heating rate $Q_{\rm total} = \rho [(z^+)^2
|z^-| + (z^-)^2 |z^+|]/4\lambda$ in the numerical solutions of the
time-independent versions of equations~(\ref{eq:Hh2}) and
(\ref{eq:Hf2}) for $g_{\rm a} = 8.0 \times 10^7 \,\mbox{cm/s}$ and for
three different values of $\lambda_0$. Going from the top of the
figure to the bottom, these values are 2500~km, 5000~km, and
$10^4$~km.  At $r\geq 2R_{\sun}$, equation~(\ref{eq:Qout})
overestimates the numerical heating rate by $\lesssim 50\%$ for the
chosen values of~$\lambda_0$. At $r< 1.6 R_{\sun}$,
equation~(\ref{eq:Qin}) overestimates the numerical heating rate by
almost an order of magnitude in the worst case ($\lambda_0 =
10^4$~km).

\begin{figure}[t]
\centerline{\includegraphics[width=8.cm]{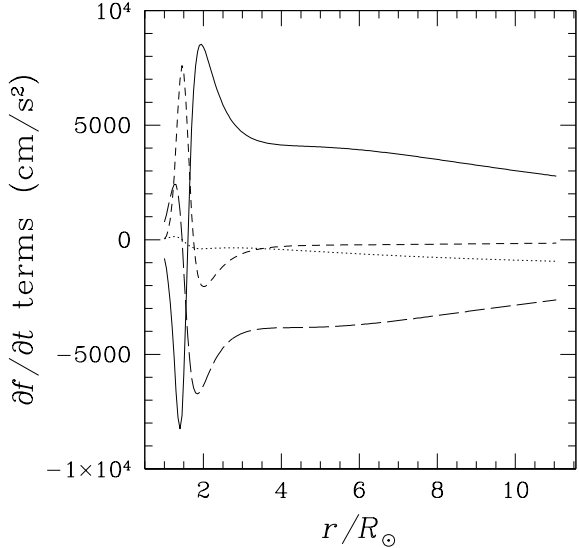}}
\caption{The different terms in equation~(\ref{eq:Hf2}), divided by $(1-\eta^{1/2})$ (see text).
The solid line is the term proportional to $\partial L/\partial s$, the long-dashed line is the nonlinear term on the right-hand side of equation~(\ref{eq:Hf2}), the dotted line is the $\partial f/\partial t$ term, and the short-dashed line is the $\partial f/\partial s$ term.
As in figure~\ref{fig:fgap}, we have taken $g_{\rm a} = 8.0 \times 10^7 \,\mbox{cm/s}$ and
  $\lambda_0 = 5000$~km.}\vspace{0.5cm}
\label{fig:fterms}
\end{figure}

In Figure~\ref{fig:fterms}, we use our time-independent solutions of
equations~(\ref{eq:Hh2}) and (\ref{eq:Hf2}) 
with $g_{\rm a} = 8.0 \times 10^7 \,\mbox{cm/s}$ and
  $\lambda_0 = 5000$~km to plot each term in
equation~(\ref{eq:Hf2}), divided by $(1- \eta^{1/2})$ to make the plot
easier to read. Although we dropped $\partial f/\partial t$ to obtain
these time-independent solutions, we estimate the approximate
magnitude of the $\partial f/\partial t$ term by multiplying our
numerical solution for $f$ by $2\pi/P$, where $P$ is the wave period,
which we take to be three hours. Our estimate of the $\partial
f/\partial t$ term is smaller than the terms that we have kept inside
the Alfv\'en critical point, but $\partial f/\partial t$ increases
relative to the dominant terms as $r$ increases towards $r_a$.  We
have compared the different terms in equation~(\ref{eq:Hf2}) at
larger~$r$, and find that our neglect of the $\partial f/\partial t$
term becomes unjustified for $r\gtrsim 20 R_{\sun}$, because the
nonlinear time scale becomes larger than the wave period (measured in
the Sun's frame) due to the increase in $\lambda$ and the decrease
in~$z^+$ as $r$ increases. Figure~\ref{fig:fterms} shows that the
$\partial f/\partial s$ term, which we have neglected in
equation~(\ref{eq:Hf2}), is negligible except in a small region near
$r=r_m$, where our solution for $f$ passes through zero. Although it
is not plotted, we have also carried out the same comparison for the
different terms in equation~(\ref{eq:Hh2}), using our time-independent
numerical solution. We find that the $\partial L/\partial s$ term that
we neglect in equation~(\ref{eq:Hh2}) is indeed smaller than the other
terms out to well beyond 20~$R_{\sun}$, because $f \ll g$. 

\begin{figure}[t]
\centerline{\includegraphics[width=8.cm]{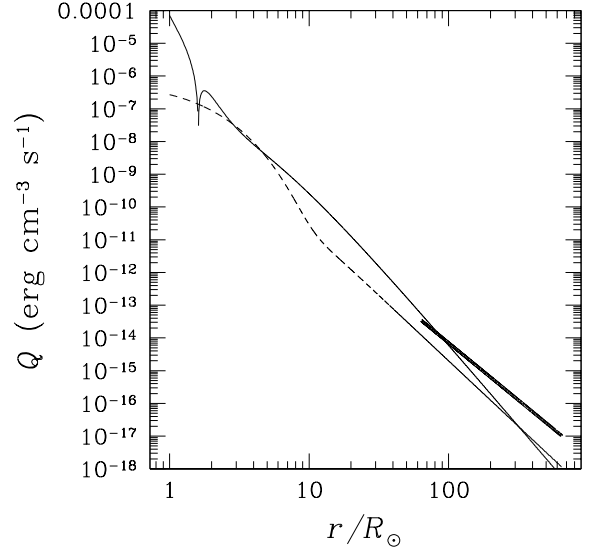}}
\caption{The thin solid line is the value of $Q$ in
  equations~(\ref{eq:Qout}) and (\ref{eq:Qin}) for $g_{\rm a} = 8.0
  \times 10^7 \,\mbox{cm/s}$ . The dashed line is the total (electron
  plus proton) heating rate in model SW2 of Allen et al. (1998). The
  thick solid line is the proton heating rate from Eq. (14) of Cranmer
  et al. (2009), multiplied by 4/3 to convert it to an approximate
  total heating rate (the proton heating rate is typically 60-90\% of
  the total heating rate in their model).}\vspace{0.5cm}
\label{fig:swh}
\end{figure}

In Figure~\ref{fig:swh}, we plot the turbulent heating rate in
equations~(\ref{eq:Qout}) and (\ref{eq:Qin}) (thin solid
line). Although the approximations leading to
equations~(\ref{eq:Qout}) and (\ref{eq:Qin}) break down at $r\gtrsim
20 R_{\sun}$, we plot equation~(\ref{eq:Qout}) out to beyond $100 R_{\sun}$ to illustrate
how the analytical formulas extrapolate to larger radii.  The heating
rate in our model depends on a single free parameter, $g_a$, which we
set equal to $8.0 \times 10^7 \; \mbox{cm/s}$ in order to roughly
match two previous empirically constrained models of plasma heating in
the fast solar wind (Allen et al.\ 1998; Cranmer et al.\ 2009). We note that
for $z^- \ll z^+$, this value of $g_a$ corresponds to a fluctuating
velocity of 200 km/s at the Alfv\'en critical point.  The dashed line
in Figure~\ref{fig:swh} is the total (electron plus proton) heating
rate $Q_{\rm A98}$ in model SW2 of Allen et al. (1998). The thick
solid line is the proton heating rate in the model of Cranmer et
al. (2009) (their Eq. 14, which applies at $r>0.3$~AU), multiplied by
4/3 to convert the proton heating rate to an approximate total heating
rate, denoted $Q_{\rm C09}$. [The proton heating rate is typically
60-90\% of the total heating rate in Cranmer et al\.'s (2009) model].

Model SW2 of Allen et al.~(1998) does not provide a tight constraint
on the heating rate, since widely different heating functions and
momentum-deposition profiles (their models SW2, SW3, and SW4) were
shown to be consistent with {\em in situ} observations of the fast
wind. We include $Q_{\rm A98}$ in Figure~\ref{fig:swh} to show that
our results for the heating rate at $2R_{\sun} < r < 7 R_{\sun}$ are reasonably close
to at least one empirically constrained model for a plausible value
of~$g_a$. On the other hand, our heating rate greatly exceeds $Q_{\rm
  A98}$ at $r<1.5R_{\sun}$. This disagreement may be due in part to
the effects described in section~\ref{sec:ts} and also to
the point illustrated in figure~\ref{fig:swh_comp}, that we
substantially overestimate the heating rate close to the Sun when
$\lambda_0> 5000$~km.  
We emphasize that although Figure~\ref{fig:swh} represents a rough first
attempt to compare our model to observational constraints on the
heating rate, this comparison must be viewed with some caution,
because the density profiles in the three models are different, and
$Q$ depends upon $\rho$ in equations~(\ref{eq:Qout}) and (\ref{eq:Qin}).

\begin{figure}[t]
\centerline{\includegraphics[width=8.cm]{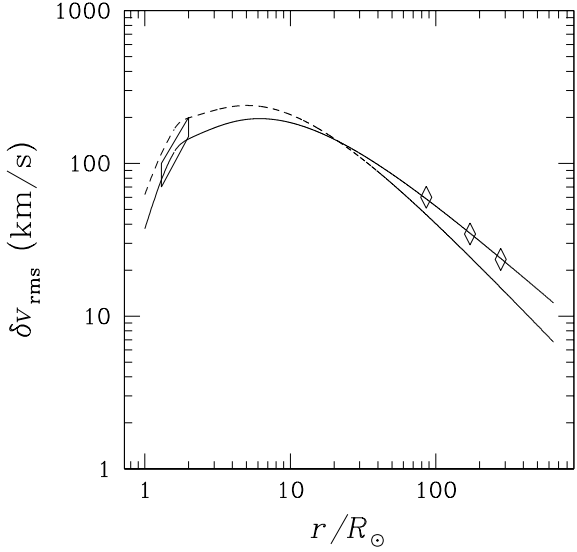}}
\caption{The dashed line is the rms value of the wave velocity in our
  low-wave-frequency model, which is approximately $z^+/2$ when $z^-
  \ll z^+$ , obtained from equations~(\ref{eq:hsolve_out}) and
  (\ref{eq:hsolve_in}) with $g_a = 8 \times 10^7 \; \mbox{cm/s}$.  The
  solid line is the rms value of the wave velocity in the extended
  model described in section~\ref{sec:extended}, obtained from
  equations~(\ref{eq:hsolve_out2}) and (\ref{eq:hsolve_in2}) with $g_a
  = 7.2 \times 10^7 \; \mbox{cm/s}$ and $\chi = 0.65$.  The open box
  is adapted from Figure 9 of CvB05, and
  represents an upper limit on the non-thermal line widths obtained by
  Esser et al. (1999) using off-limb UVCS measurements (Esser et
  al. 1999). The open diamonds are from {\em in situ} Helios and Ulysses
  measurements (Bavassano et al. 2000).  }\vspace{0.5cm}
\label{fig:swv_both}
\end{figure}

In Figure~\ref{fig:swv_both}, we plot the rms amplitude of the
fluctuating velocity in our low-wave-frequency model as a function
of~$r$ (dashed line), using equations~(\ref{eq:hsolve_out}) and
(\ref{eq:hsolve_in}) and the relation $\delta v_{\rm rms} \simeq
\langle (z^+)^2\rangle ^{1/2}/2$, which is valid when $z^- \ll
z^+$. We also plot observed values and upper limits from remote 
UVCS observations of the corona (Esser et al. 1999; CvB05) and {\em in situ} measurements from Helios and
Ulysses (Bavassano et al. 2000). (The solid line in this figure is
from the extended model described in section~\ref{sec:extended}.)  For
$g_a = 8 \times 10^7\; \mbox{cm/s}$, our low-wave-frequency model
over-predicts the rms amplitude of velocity fluctuations at $1.5
R_{\sun} < r < 2R_{\sun}$. As noted above, the approximations leading
to equations~(\ref{eq:hsolve_out}) and (\ref{eq:hsolve_in}) break down
at $r\gtrsim 20 R_{\sun}$. Nevertheless, we plot
equation~(\ref{eq:hsolve_out}) out to larger~$r$ to illustrate how the
analytical formulas extrapolate to larger radii. This extrapolation
shows that for $g_a = 8 \times 10^7 \; \mbox{cm/s}$,
equations~(\ref{eq:hsolve_out}) and (\ref{eq:hsolve_in}) under-predict
$\delta v_{\rm rms}$ at~$r>70 R_{\sun}$.

\section{Extended Model Accounting for Shorter-Period Waves}
\label{sec:extended} 

In section~\ref{sec:lwfm}, we took the waves launched by the Sun to
have periods of hours and longer. On the other hand, CvB05 suggested
that a significant fraction of the outward wave power near the Sun is
in waves with periods of 1~minute to 1~hour.\footnote{We note,
  however, that VV07 considered two different frequency spectra for
  the~$z^+$ waves at the coronal base: a flat spectrum in which the
  energy is mostly in waves with periods~$P$ of $\sim 10$~minutes (the
  small-$P$ cutoff in their spectrum), and a steep spectrum in which
  most of the energy is in the longest-period waves in their model,
  with $P\sim 10^3$~hours. With the steep spectrum, their model agreed
  with {\em in situ} measurements of the $z^+$ and $z^-$ amplitudes, but
  with the flat spectrum their model over(under)-estimated the $z^+$
  ($z^-$) energy relative to the {\em in situ} measurements. The VV07
  results thus favor a picture in which the Sun launches primarily
  long-period waves.} In this section, we extend our model to account
for such shorter-period waves in an approximate manner. The
approximations we use are very simplistic, but are not limited to
$r\lesssim 20 R_{\sun}$.

In the spirit of our simple, phenomenological modeling, we begin by
replacing each ${\bf g}$ and ${\bf f}$ in equations~(\ref{eq:Hh2}) and
(\ref{eq:Hf2}) with the (time-independent) rms values $\langle
g^2\rangle^{1/2}$ and $\langle f^2\rangle^{1/2}$, respectively. 
As shown by Velli (1993) and others, wave reflection becomes less
efficient as the wave period decreases below $\sim 1 $~hour, since higher-frequency
waves with shorter parallel wavelengths are less affected by the
radial gradient in the Alfv\'en speed. To account for this effect,
 we set
\begin{equation}
\frac{\partial L}{\partial s} \rightarrow \chi \frac{\partial L}{\partial s}
\label{eq:defchi} 
\end{equation} 
in equations~(\ref{eq:Hh2}) and (\ref{eq:Hf2}), where $\chi$ is a
dimensionless constant. In the low-wave-frequency limit of
section~\ref{sec:lwfm}, $\chi \rightarrow 1$. To model the mix of wave
frequencies launched from the Sun, we leave $\chi$ as a free parameter
between 0 and~1.  We take $\langle g^2\rangle^{1/2}$ and $\langle
f^2\rangle^{1/2}$  to have scale lengths in
the direction of ${\bf B}_0$ of order~$r$. We continue to assume that
$f \ll g$, and set $(\eta - 1) \partial L/\partial s \rightarrow
|(\eta - 1) \partial L/\partial s|$ in equation~(\ref{eq:Hf2}) to
maintain the positivity of $\langle f^2\rangle$, so that
equation~(\ref{eq:Hf2}) becomes
\begin{equation} 
\langle f^2\rangle^{1/2} = \frac{ 2\chi\lambda |\eta - 1| U}{\eta^{1/4}} \;\left|\frac{\partial L}{\partial s}\right|.
\label{eq:fsolve2} 
\end{equation} 
Dropping the third term on the left-hand side of equation~(\ref{eq:Hh2})  and using 
equation~(\ref{eq:fsolve2}), we obtain
\begin{equation}
\frac{\partial }{\partial s} \langle g^2 \rangle = - 2\chi \langle g^2 \rangle  \left| \frac{\partial L}{\partial s}\right|,
\label{eq:hdiffeq2} 
\end{equation} 
which we integrate to obtain
\begin{equation}
\langle g^2 \rangle ^{1/2} = g_a \left(\frac{v_{\rm A}}{v_{{\rm A}a}}\right)^{\chi/2}
\label{eq:hsolve_out2} 
\end{equation} 
for $r> r_m$ and
\begin{equation}
\langle g^2 \rangle ^{1/2} = g_a \left(\frac{v_{{\rm A}m}^2}{v_{{\rm A}a}v_{\rm A}}\right)^{\chi/2}
\label{eq:hsolve_in2} 
\end{equation} 
for $r< r_m$.
From equation~(\ref{eq:defQ}), we then find that
\begin{equation}
Q = \frac{\chi\rho \eta^{1/2} g_a^2 U}{4v_{\rm A} (1 + \eta^{1/2})} \left(\frac{v_{\rm A}}{v_{{\rm A}a}}\right)^\chi\; \left|\frac{dv_{\rm A}}{dr}\right|
\label{eq:Qout2} 
\end{equation} 
for $r>r_m$, and
\begin{equation}
Q = \frac{\chi\rho \eta^{1/2} g_a^2 U}{4v_{\rm A} (1 + \eta^{1/2})}\left(\frac{v_{{\rm A}m}^2}{v_{{\rm A}a} v_{\rm A}}\right)^\chi \frac{dv_{\rm A}}{dr}
\label{eq:Qin2} 
\end{equation} 
for $r< r_m$. The results of our low-frequency model in section~\ref{sec:lwfm}  are recovered by setting
$\chi =1$ in equations~(\ref{eq:fsolve2}) through (\ref{eq:Qin2}). 

\begin{figure}[t]
\centerline{\includegraphics[width=8.cm]{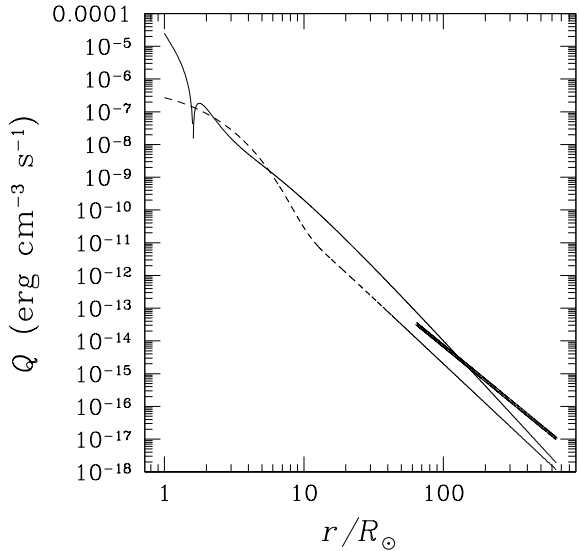}}
\caption{The solid line is the value of $Q$ in equations~(\ref{eq:Qout2}) and (\ref{eq:Qin2}) 
for $g_{\rm a} = 7.2 \times 10^7 \,\mbox{cm/s}$ and $\chi = 0.65$. The dashed line is the total (electron plus proton) heating rate in model SW2 of Allen et al. (1998). The dotted line is the proton heating rate from Eq. (14) of Cranmer et al. (2009), which is typically 60-90\% of the total heating rate in their model.
}\vspace{0.5cm}
\label{fig:swh_ref}
\end{figure}

\begin{figure}[t]
\centerline{\includegraphics[width=8.cm]{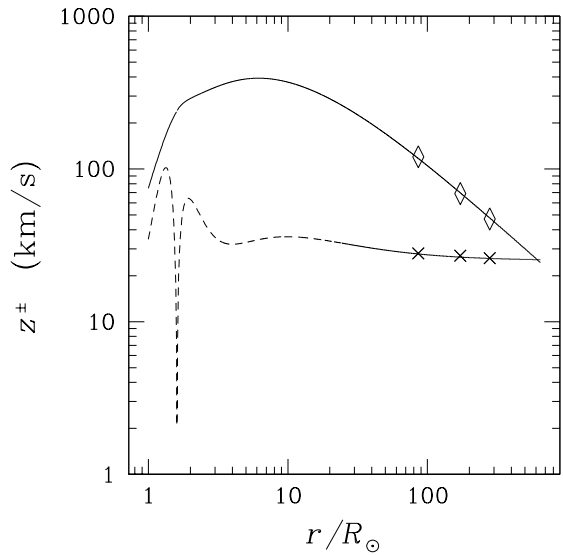}}
\caption{The solid line is the rms amplitude of $z^+$ for $g_a = 7.2
  \times 10^7 \; \mbox{cm/s}$ and $\chi = 0.65$ from
  equations~(\ref{eq:hsolve_out2}) and (\ref{eq:hsolve_in2}).  The
  dashed line is the rms amplitude of $z^-$ from
  equation~(\ref{eq:fsolve}) for 
$\lambda_0 = 1.6\times 10^4\;  \mbox{km}$ 
and the same value of~$\chi$.  The diamonds and crosses
  are the rms values of $z^+$ and $z^-$ , respectively, from Bavassano
  et al. (2000).  }\vspace{0.5cm}
\label{fig:swzpm_ref}
\end{figure}

When  $z^- \ll z^+$, the rms amplitude of the fluctuating velocity  $\delta v_{\rm rms}$
is approximately~$\langle (z^+)^2\rangle ^{1/2}/2$, and there
are two free parameters that determine $\delta v_{\rm rms}$: $\,g_a$
and $\chi$. We set $g_a = 7.2 \times 10^7\;\mbox{cm/s}$ and $\chi =
0.65$ in order to match the observational constraints on the the
fluctuating velocity, shown in Figure~\ref{fig:swv_both}. The solid
line in Figure~\ref{fig:swv_both} is the resulting value of $\delta
v_{\rm rms}$ obtained from equations~(\ref{eq:hsolve_out2}) and
(\ref{eq:hsolve_in2}). In Figure~\ref{fig:swh_ref} we plot the heating
rate from equations~(\ref{eq:Qout2}) and (\ref{eq:Qin2}) for these
values of $g_a$ and $\chi$, along with the empirically constrained
heating rates $Q_{\rm A98}$ and $Q_{\rm C09}$ from the models of Allen
et al.~(1998) and Cranmer et~al.~(2009) described in
section~\ref{sec:lwfm}.  We plot the rms amplitudes of $z^+$ and $z^-$
resulting from equations~(\ref{eq:fsolve2}), (\ref{eq:hsolve_out2}),
and (\ref{eq:hsolve_in2}) in Figure~\ref{fig:swzpm_ref}, as well as
{\em in situ} measurements from Helios and Ulysses (Bavassano et
al. 2000).  The rms amplitude of $z^-$ that results from
equation~(\ref{eq:fsolve2}) depends upon both $\chi$ and
$\lambda_0$. To match the {\em in situ} data, we set $\lambda_0 =
1.6\times 10^4$~km in this figure, keeping $\chi = 0.65$ as in
Figures~\ref{fig:swv_both} and~\ref{fig:swh_ref}.
As Figure~\ref{fig:swzpm_ref}  shows,
the approximation $z^- \ll z^+$ fails at~$r\gtrsim 1$~AU.

Overall, the model matches the observations and previous models fairly
well. However, as in our low-wave-frequency model, the heating rate in
our extended model is much larger than $Q_{\rm A98}$ at $r<
1.5R_{\sun}$.  This discrepancy may arise in part for the reasons
discussed in section~\ref{sec:ts}. In addition, our extended model is
likely similar to our low-frequency model in significantly
overestimating~$Q$ close to the Sun when $\lambda_0 \geq 5000$~km
(Figure~\ref{fig:swh_comp}). A second difference between the heating
rates plotted in Figure~\ref{fig:swh_ref} is that $Q$ becomes smaller
than $Q_{\rm C09}$ at $r> 1$~AU. This discrepancy may arise because of
the increasing importance at large~$r$ of sources of inward waves
(such as velocity shear) that are not included in our model.  We note,
however, that some caution is warranted when comparing our heating
rate to $Q_{\rm A98}$ and $Q_{\rm C09}$, because the density profiles
in the three models are different, and $Q$ depends upon $\rho$ in
equations~(\ref{eq:Qout2}) and (\ref{eq:Qin2}).

Compared to our model for low-frequency waves in
section~\ref{sec:lwfm}, our extended model accounting for
shorter-period waves does a notably better job of reproducing the
observational constraints on the velocity fluctuations in the solar
wind. This is not surprising, given that there is an additional free
parameter in the extended model and the approximations underlying our
low-wave-frequency model break down at $r\gtrsim 20 R_{\sun}$. There
is, however, an additional reason for the closer agreement between the
extended model and the data. As can be seen in
Figure~\ref{fig:swv_both}, the value of $\delta v_{\rm rms}$ in our
low-wave-frequency model drops off more rapidly with radius than does
the data. When the efficiency of reflection is reduced, the rate at
which the outward wave energy cascades and dissipates is reduced, and
so $| (d/dr) \delta v_{\rm rms}|$ decreases.

\section{The Energy Cascade Time for $z^+$ Waves}
\label{sec:ts} 

\begin{figure}[t]
\centerline{\includegraphics[width=8.cm]{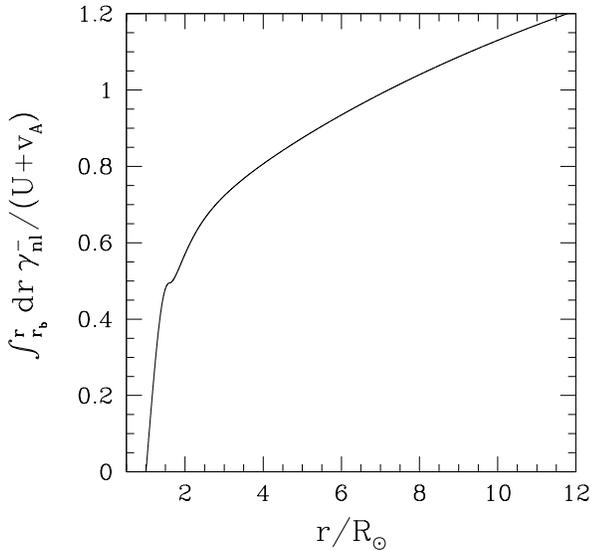}}
\caption{The time integral of the shearing rate experienced by a $z^+$
  wave packet between the time it leaves the coronal base at $r=r_b
  \simeq 1 R_{\sun}$ and the time it reaches radius~$r$, from
  equation~(\ref{eq:Gamma2}) with $\chi=1$.}\vspace{0.5cm}
\label{fig:sw_gammatnl}
\end{figure}

If the Sun launches an outward-propagating ($z^+$) Alfv\'en wave
packet with a perpendicular scale length~$\lambda_0$ at the coronal
base, with $\lambda_0$ greatly exceeding the dissipation scale, then
this wave packet must propagate some distance into the corona before its
energy cascades to small perpendicular scales and dissipates. We
estimate this distance using approximations and results from our
analytical models.  As in equation~(\ref{eq:Hh2}), the rate at which
$z^-$ waves shear $z^+$ waves is
\begin{equation}
\gamma_{\rm nl}^- = \frac{|z^-|}{2\lambda}.
\label{eq:gammanl} 
\end{equation} 
The quantity
\begin{equation}
\Gamma = \int _{r_b}^r \frac{|z^-| dr_1}{2\lambda(U+v_{\rm A})}.
\label{eq:Gamma} 
\end{equation} 
is the time integral of~$\gamma_{\rm nl}^-[{\bf r}(t), t]$ evaluated at a position~${\bf r}(t)$ that moves outward with the $z^+$ wave packet at speed~$U+v_{\rm A}$, starting from the 
time the wave packet leaves the coronal base at $r=r_b \simeq
1 R_{\sun}$ until the time the wave packet reaches radius~$r$.
Roughly speaking, the $z^+$ wave packet must propagate out to a distance at
which~$\Gamma \gtrsim 1$ before its energy cascades and dissipates.
We evaluate the right-hand side of equation~(\ref{eq:Gamma}) in our
low-frequency model using equation~(\ref{eq:fsolve}). To evaluate
equation~(\ref{eq:Gamma}) using our extended-model results, we replace
$z^-$ with $\langle (z^-)^2\rangle^{1/2}$ and use
equation~(\ref{eq:fsolve2}). The integral in equation~(\ref{eq:Gamma}) 
can then be evaluated analytically to yield 
\begin{equation}
\Gamma = \left\{
\begin{array}{ll}
(\chi/2)\ln(v_{\rm A}/v_{{\rm A}b}) & \mbox{ if $r<r_m$} \vspace{0.4cm} \\
(\chi/2)\ln[v_{{\rm A}m}^2/(v_{{\rm A}b}v_{\rm A})] & \mbox{ if $r>r_m$}
\end{array}
\right. ,
\label{eq:Gamma2} 
\end{equation}  
where the result for our low-frequency model is obtained by setting
$\chi=1$.  The value of $\Gamma$ is plotted in
Figure~\ref{fig:sw_gammatnl}, assuming $\chi=1$ and using the profiles
for $n$, $B_0$, and~$U$ given in equations~(\ref{eq:n}) through
(\ref{eq:defU}). The condition $\Gamma > 1$ only holds for $r\gtrsim 7
R_{\sun}$, suggesting that the turbulent heating rate may be
smaller than in our model within a few~$R_{\sun}$ of the
solar surface.  We note, however, that the reduction in~$Q$ due to the
condition~$\Gamma <1$ may itself be mitigated by a further complication,
that turbulence within the chromosphere [where $z^+ \sim z^-$ (CvB05)]
may lead to the launching of Alfv\'en waves into the corona with a
broad range of perpendicular length scales, extending to very small
values (Chandran~2008). Further work is needed to explore how these effects modify the
radial profile of the turbulent heating rate.

In one sense, Figure~\ref{fig:sw_gammatnl} provides a rough
consistency check on models in which the solar wind is heated and
accelerated by low-frequency Alfv\'en-wave turbulence. If $\Gamma$
were $\ll 1$ at $r=r_a$, then only a tiny fraction of the
Alfv\'en-wave energy launched from the base of the Sun would dissipate
in the region where most of the heating and acceleration of the solar
wind takes place. A small fraction of the Alfv\'en-wave energy would
be unable to generate the solar wind, because the Alfv\'en wave energy
flux measured {\em in situ} at $r>0.3$~AU is much smaller than the
total solar-wind energy flux, even allowing for loss of Alfv\'en-wave
energy due to work done on the flow by the ponderomotive force.  On
the other hand, if $\Gamma$ rose from 0 at $r=r_b$ to a value $\gg 1$
at say $r=7R_{\sun}$, then most of the wave energy would dissipate
very close to the Sun (within the radius at which $\Gamma$ reached a
value of a few). If the heating rate were to fall too rapidly
with~$r$, then the heating profile would become inconsistent with UVCS
observations that show ion temperatures staying flat or increasing
with radius out to at least~$3.5 R_{\sun}$ despite adiabatic cooling
(Kohl et~al~1998; Antonucci, Dodero, \& Giordano 2000).

\section{Conclusion}

In this paper, we consider Alfv\'en-wave reflection and turbulent
heating in the solar wind using the Heinemann-Olbert equations, which
describe non-compressive MHD fluctuations in an inhomogeneous medium
with a background flow parallel to the background magnetic field. We
approximate the nonlinear terms in these equations using a simple
phenomenological model from D02.  Wave reflection plays an essential role
in our calculation of the turbulent heating rate, because only
outward-propagating waves are generated by the Sun, and turbulent
interactions arise only from interactions between oppositely
propagating waves (for the low-frequency, non-compressive Alfv\'en
waves that we consider).

In section~\ref{sec:lwfm}, we restrict our attention to waves with
periods of hours and longer. Our focus on low-frequency waves in this
section is motivated by {\em in situ} observations showing that most
of the wave power is at periods of hours in the spacecraft frame. We
assume that the energy in inward-propagating waves is much less than
the energy in outward-propagating waves, i.e., $z^- \ll z^+$, an
approximation that is appropriate for coronal holes and the fast solar
wind at $r< 1$~AU. We also assume that the wave period is longer than
the energy cascade time for Sunward-propagating waves,~$\sim
\lambda/z^+$.  With the use of these approximations, we solve our
model equations analytically to obtain expressions for the rms
amplitudes of outward and inward waves as a function of~$r$. We also
obtain an analytical expression for the radial profile of the
turbulent heating rate.  There is one free parameter in the heating
rate in our low-wave-frequency model: the amplitude of the
outward-propagating waves at some fixed reference point. We take this
reference point to be the Alfv\'en critical point~$r=r_A$, which is at
$11.1 R_{\sun}$ in our model solar wind.  Interestingly, as in the
model of D02, the heating rate does not depend upon the choice of the
dominant perpendicular length scale of the turbulence,
$\lambda$. Ordinarily, the heating rate is inversely proportional to
$\lambda$. However, in our model, the amplitude of the
inward-propagating waves is proportional to $\lambda$, because small
$\lambda$ leads to rapid turbulent dissipation of the inward waves. As
a result, $Q\simeq \rho\langle (z^+)^2\rangle \langle
(z^-)^2\rangle^{1/2}/4\lambda$ becomes independent of~$\lambda$, as in
equation~(5) of~D02. Our model can be thought of as generalizing the
phenomenological model of D02 by accounting for the solar wind
velocity, so that the model can be applied all the way from the
coronal base out past the Alfv\'en critical point.
 
As discussed in section~\ref{sec:lwfm}, one of the approximations in
our low-wave-frequency model --- that the energy cascade time of
Sunward-propagating waves is shorter than the wave period measured in
the Sun's frame --- breaks down for $r\gtrsim 20 R_{\sun}$ for a wave
period of $\sim 3$~hours. In addition, our low-wave-frequency model
does not account for waves with periods much shorter than one hour,
which may make a significant contribution to the outward wave flux
from the Sun (CvB05; but see VV07).  To overcome these limitations, we
develop an ``extended'' model in section~\ref{sec:extended} that
approximately accounts for waves with shorter periods and does not
require the nonlinear time scale to be shorter than the wave period in
the reference frame of the Sun. The main additional ingredient in this
extended model is a new free parameter, $\chi$, that models the
decrease in the efficiency of wave reflection at shorter wave
periods. For appropriate choices of the two free parameters in the
model ($\chi = 0.65$ and $g_a = 7.2 \times 10^7\;\mbox{cm/s}$), our
extended model is successful at matching observations of velocity
fluctuations and the Els\"asser fields.

Both of the models we have developed provide analytical expressions
for the turbulent heating rate that can be used to incorporate
Alfv\'en-wave reflection and turbulent heating into fluid models of
the solar wind, both inside and outside the Alfv\'en critical
point. We conclude by summarizing the relative advantages and
disadvantages of the two models:
\begin{enumerate}
\item The low-wave-frequency model described in section~\ref{sec:lwfm}
  involves a simplistic phenomenological approximation of only the
  nonlinear terms in the basic equations.  The heating rate in the model involves
  only a single free parameter, $g_a$, which determines the amplitude
  of the outward waves at the Alfv\'en critical point. On the other
  hand, one of the assumptions of the model breaks down at $r\gtrsim
  20 R_{\sun}$, where the nonlinear time scale becomes longer than the
  wave period measured in the frame of the Sun (for wave periods of
  $\sim 3$~hours). In addition, the model does not account for waves
  with periods much shorter than 1~hour, which may contribute
  significantly to the outward wave flux (CvB05).
\item The extended model described in section~\ref{sec:extended} does
  a better job of matching observational constraints on the wave
  amplitudes, accounts (in a very approximate way) for both
  long-period waves and waves with periods much less than one hour,
  and can be applied all the way out to 1~AU. On the other hand, this
  model involves simple phenomenological modeling of both the
  nonlinear terms and the linear reflection term. The heating rate in
  the model also involves two free parameters instead of one.
\end{enumerate}

\acknowledgements We thank the anonymous referee for a very helpful
report. This work was supported in part by NSF Grant ATM-0851005, by
DOE Grant DE-FG02-07-ER46372, by NSF-DOE Grant AST-0613622, and by
NASA Grants NNX07AP65G and NNX08AH52G.

\references

Allen, L. A., Habbal, S. R., \& Hu, Y. Q. 1998, J. Geophys. Res., 103, 6551

Antonucci, E., Dodero, M. A., \& Giordano, S. 2000, Solar Phys., 197, 115

Barnes, A. (1981), Turbulence and Dissipation in the Solar Wind, in Solar Wind Four, edited by H. Rosenbauer, p. 326, Max Planck Institut für Aeronomie Report No. MPAE-W-100-81-31, Katlenburg-Lindau.

Bavassano, B., Pietropaolo, E., \& Bruno, R. 2000, J. Geophys. Res., 105, 15959

Belcher, J. W., \& Davis, L. 1971, J. Geophys. Res., 76, 3534

Belcher, J. W., Davis, L., \&  Smith, E. J. 1969, J. Geophys. Res., 74, 2302

Breech, B., Matthaeus, W. H., Minnie, J., Bieber, J. W., Oughton, S., Smith, C. W., \& Isenberg, P. A. 2008, J. Geophys. Res., 113, A08105

Bruno, R., \& Carbone, V. 2005, Living Rev. Sol. Phys., 2, 4

Chandran, B. D. G. 2008, ApJ, 685, 646

Chandran, B. D. G., Quataert, E., Howes, G. G., Hollweg, J. V., \& Dorland, W. 2009, ApJ, 701, 652 

Coleman, P. J. 1968, ApJ, 153, 371639, 1177 

Cranmer, S. R., Matthaeus, W. H., Breech, B. A., \& Kasper, J. C. 2009, ApJ, 702, 1604

Cranmer, S. R. \& van Ballegooijen, A. A. 2005, ApJS, 156, 265 (CvB05)

Cranmer, S. R., van Ballegooijen, A. A., \& Edgar, R. J. 2007, ApJS, 171, 520

Del Zanna, L., Velli, M., \& Londrillo, P. 2001, A\&A, 367, 705

Dmitruk, P., Milano, L. J., \& Matthaeus, W. H. 2001, ApJ, 548, 482

Dmitruk, P., Matthaeus, W. H., Milano, L. J., Oughton, S., Zank, G. P., \& 
Mullan, D. J. 2002, ApJ, 575, 571  (``D02'')

Dmitruk, P., \& Matthaeus, W. H. 2003, ApJ, 597, 1097

Esser, R., \& Sasselov, D. 1999, ApJ, 521, L145

Feldman, W. C., Habbal, S. R., Hoogeveen, G., \& Wang, Y.-M. 1997, J. Geophys. Res., 102, 26905

Galeev, A. A., \& Oraevskii, V. N. 1963, Sov. Phys.---Dokl., 7, 988

Goldstein, M. L., Roberts, D. A., \& Matthaeus, W. H. 1995, ARAA, 33, 283

Grappin, R., \& Velli, M. 1991, Ann. Geophys., 9,  416 

Grappin, R., Velli, M., \& Mangeney, A. 1993, Phys. Rev. Lett., 70, 2190

Heinemann, M., \& Olbert, S. 1980, J. Geophys. Res., 85, 1311

Hollweg, J.V. (1983), Coronal heating by waves, in Solar Wind Five, edited by M. Neugebauer, p. 3, NASA Conf. Publ. 2280, Washington DC.

Hollweg, J. V. 1986, J. Geophys. Res., 91, 4111

Hollweg, J. V., \& Isenberg, P. A. 2002, J. Geophys. Res., 107, 1

Hollweg, J.V., \&  Isenberg, P. A. 2007,  J. Geophys. Res., 112, A08102 

Hollweg, J. V., \& W. Johnson 1988, J. Geophys. Res., 93, 9547

Kadomtsev, B. B. \&  Pogutse, O. P. 1974,  Sov. JETP, 38, 283

Marsch, E., \& Tu, C.-Y. 1996, Ann. Geophys. 14, 259

Matthaeus, W. H., Zank, G. P., Oughton, S. , Mullan, D. J.,  \& Dmitruk, P. 1999, ApJ, 523, L93

Roberts, D. A., Goldstein, M. L., Matthaeus, W. H., \& Ghosh, S. 1992,
J. Geophys. Res., 97, 17115

Schekochihin, A. A., Cowley, S. C., Dorland, W., Hammett, G. W., Howes, G. G., Quataert, E., \& Tatsuno, T. 2009, ApJS, 182, 310

Spruit, H. 1981, ``Magnetic flux tubes,'' in {\em The Sun as a Star}, ed. S. Jordan, p. 385, NASA SP-450, Washington DC.

Strauss, H. R. 1976, Phys. Fluids, 19, 134

Tu, C.-Y. 1987, Solar Phys., 109, 149

Tu, C.-Y. 1988, J. Geophys. Res., 93, 7

Tu, C.-Y., \& Marsch, E. 1995, Sp. Sci. Rev., 73, 1

Tu, C.-Y., Pu, Z.-Y, \& Wei, F.-S. 1984, J. Geophys. Res., 89, 9695

Velli, M. 1993, Astron. Astrophys., 270, 304

Velli, M., Grappin, R., \& Mangeney, A. 1989, Phys. Rev. Lett., 63, 1807

Velli, M., Grappin, R., \& Mangeney, A. 1990, Comp. Phys. Comm., 59,  153

Verdini, A., \& Velli, M. 2007, ApJ, 662, 669  (VV07)

Verdini, A., Velli, M., \& Buchlin, E., 2009, ApJL, 700, 39

Vi\~nas, A. F., \& Goldstein, M. L. 1991, J. Plasma Phys., 46, 129

Zank, G. P. \& Matthaeus, W. H. 1992, J. Plasm. Phys., 48, 85

\end{document}